\documentclass[12pt]{article}
\usepackage[dvips]{graphicx}
\usepackage{latexsym}
\def\slashchar#1{\setbox0=\hbox{$#1$}           
   \dimen0=\wd0                                 
   \setbox1=\hbox{/} \dimen1=\wd1               
   \ifdim\dimen0>\dimen1                        
      \rlap{\hbox to \dimen0{\hfil/\hfil}}      
      #1                                        
   \else                                        
      \rlap{\hbox to \dimen1{\hfil$#1$\hfil}}   
      /                                         
   \fi}                                         %

\newcommand{\bc}{\begin{center}}
\newcommand{\ec}{\end{center}}
\newcommand{\be}{\begin{equation}}
\newcommand{\ee}{\end{equation}}
\newcommand{\bea}{\begin{eqnarray}}
\newcommand{\eea}{\end{eqnarray}}
\newcommand{\ba}{\begin{eqnarray}}
\newcommand{\ea}{\end{eqnarray}}
\newcommand{\brr}{\begin{array}}
\newcommand{\err}{\end{array}}

\newcommand{\simge}{\ \lower-
1.2pt\vbox{\hbox{\rlap{$>$}\lower5pt
\vbox{\hbox{$\sim$}}}}\ }

\begin{document}
\pagestyle{empty} 
\vspace{-0.6in}
\begin{flushright}
BUHEP-00-22\\
\end{flushright}
\vskip 1.5in

\centerline{\large {\bf{Schwinger Model with the Overlap-Dirac Operator: exact}}} 
\centerline{\large {\bf{results versus a physics motivated approximation}}}
\vskip 0.4cm
\centerline{\bf{Leonardo Giusti, Christian Hoelbling, Claudio Rebbi}} 
\vskip 0.8cm
\centerline{Boston University - Department of Physics} 
\centerline{590 Commonwealth Avenue, Boston MA 02215}
\centerline{USA}
\vskip 0.3cm
\centerline{e-mail:     lgiusti@bu.edu}
\centerline{$\;\;\;\;\;\;\;$ hch@bu.edu}
\centerline{$\;\;\;\;\;\;\;\;\;$ rebbi@bu.edu}
\vskip 1.0in
\begin{abstract}
\noindent We propose new techniques 
for the numerical implementation of the
overlap-Dirac operator, which exploit the physical properties of the
underlying theory to avoid nested algorithms. We test these procedures 
in the two-dimensional Schwinger model and the results are 
very promising. These techniques can be directly applied to 
QCD simulations. We also present a detailed computation of the 
spectrum and the chiral properties of the Schwinger model 
in the overlap lattice formulation. 
\end{abstract}
\vfill
\pagestyle{empty}\clearpage
\setcounter{page}{1}
\pagestyle{plain}
\newpage 
\pagestyle{plain} \setcounter{page}{1}

\newpage

\section{Introduction}

The last few years witnessed a major breakthrough in the lattice
regularization of Fermi fields.  The closely related domain 
wall \cite{DK} and overlap \cite{HN,neub1} formulations of lattice 
fermions provide a definition of a lattice Dirac
operator $D$ which avoids the doubling problem and preserves the
relevant symmetries of the continuum theory, most notably chiral
symmetry in the limit of vanishing fermion mass.  Unfortunately, this
welcome development has come at a price: the numerical calculation of
the matrix elements of the propagator $D^{-1}$ and the inclusion of
${\rm Det}(D)$ in the measure entail a substantially increased
computational burden, which severely constrains the maximum lattice
size for viable simulations.  Many efforts have recently been devoted
to finding more efficient ways to perform these computational tasks
\cite{A}.  The problem has generally been approached from a
numerical analysis point of view, looking for approximations that
allows one to use sparse matrix techniques, although $D$ itself is not
sparse.  In this article we would like to advocate and explore a
different approach, where the approximation is based on the
physics and proceeds through a projection over a
subspace, which has a substantially reduced number of degrees freedom
but still captures the relevant long range properties.  
Our approximation consists in taking the matrix elements
of the Wilson Dirac operator in a subspace consisting of the long
range modes which we expect to dominate the low energy
properties of the theory and in constructing the overlap Dirac
operator numerically, but without further approximations, in this
subspace.  In the complement, the Dirac operator is approximated by
its free form.  The projection over long range modes is done via a
Fourier transform after gauge fixing.  We also studied a gauge
invariant method of projection.  We will compare the results obtained
with our approximation in a simplified model with those of an exact
calculation, finding satisfactory agreement.  It is our hope that the
approximation may offer a new way to implement the overlap formulation
in four-dimensional QCD.

We will focus on the overlap Dirac operator and we will apply our
approximation procedure to the two-dimensional Schwinger model.  We
use the overlap formulation because it provides an exact framework
for the new lattice fermions.  The domain wall formulation becomes
exact in the limit of infinite extent in the extra dimension, in which
case it becomes equivalent to the overlap formulation up to
corrections that vanish with the lattice spacing.  With finite extent
in the extra dimension, the domain wall formulation can be viewed as a
numerical approximation to the exact result.  Since one of our goals
is to compare the results of our approximation to exact results, the
overlap formulation is better indicated.  We study the Schwinger
model because the system is simple enough that we will be able to
calculate $D^{-1}$ and ${\rm Det}(D)$ exactly as well as in our
proposed scheme of approximation.  The results of the exact
calculations are, we believe, interesting per se, because they are
more extensive than what, to the best of our knowledge, has been done
up to now and validate in an impressive manner the advantages of the
new formulation of lattice fermions.

The plan of this paper is as follows.  In the next section we will
briefly review known properties of the Schwinger model in the continuum 
and in the lattice overlap formulation. In Section~\ref{sec:exact} we will present the
results of a numerical simulation where propagator and determinant of
the overlap Dirac operator are calculated exactly.  In Section~\ref{sec:approx} we
will introduce the proposed scheme of approximation, which is based on
the projection over a reduced number of degrees of freedom.  We will
compare the results obtained with this approximation with those of the
exact calculation.  In Section~\ref{sec:determinant} we will describe an
approximation to the determinant of $D$, based on a coarsening of the
lattice and compare results obtained with the full and approximated
determinants.  In the last section we will present a few words of
conclusion.

\section{The Schwinger Model and the Overlap Formulation}
\label{sec:schw_model}

We consider the Schwinger Model in (two-dimensional) Euclidean space-time
and we allow for $N_f$ degenerate flavors.  The model is defined by the action
\be
S = \int d^2 x \left[\frac{1}{4} F_{\mu\nu} F_{\mu\nu} +     
\sum_{i=1,N_f}\bar\psi_i(D_\mu \gamma_\mu  + m )\psi_i  
\right]
\ee 
where $\psi_i$, $\bar\psi_i$ are two components spinors, $A_\mu$
is the $U(1)$ gauge potential,  $D_\mu=\partial_\mu + ig A_\mu$ 
and $F_{\mu\nu}=\partial_\mu A_\nu - \partial_\nu A_\mu$.
We will use the following two-dimensional representation 
of $\gamma$-matrices
\be
\gamma_1 = \sigma_1\; , \qquad \gamma_2 = \sigma_2\; , \qquad 
\gamma_5 = -i \gamma_1\gamma_2 = \sigma_3
\ee 
where $\sigma_i$ are the Pauli matrices.

The topological charge of a classical gauge configurations is 
\be
Q(A) \equiv \frac{g}{4\pi} \int d^2x\, \epsilon_{\mu\nu} F_{\mu\nu}(x)
\ee
while the index of the Dirac operator 
is given by the difference between the numbers of positive ($n_+$) 
and negative ($n_-$) 
chirality zero modes 
\be
\mbox{index}(A) \equiv n_- - n_+
\ee
The Atiyah-Singer theorem states that
\be
Q(A) = \mbox{index}(A) \equiv n_- - n_+
\ee
Moreover in two dimensions the vanishing theorem ensures that \cite{VT} 
\ba
n_+ \neq 0 & \rightarrow & n_-=0\nonumber\\
n_- \neq 0 & \rightarrow & n_+=0
\ea
At the quantum level the Schwinger model does not require infinite
renormalization: $g$ and $m$ are finite bare parameters. 
In the massless limit the model can can be solved exactly 
\cite{schwinger}.
In the following we will analyze the systems with $N_f \leq 2$. 
Information for generic $N_f$ can be found in~\cite{gatseil}.

\subsection{The Schwinger model with $N_f=1$.}
\label{sec:schw_model_sub1}
The classical theory has a vector $U(1)_V$ symmetry and a softly 
broken axial symmetry $U(1)_A$, which at the quantum level is also broken 
by the anomaly. The Ward Identities (WI) for the corresponding quantum 
theory are
\be\label{eq:WIs}
\partial_\mu V_\mu = 0 \; , \qquad \partial_\mu A_\mu = 
\frac{g}{2\pi}\epsilon_{\mu\nu} F_{\mu\nu} + 2 m P 
\ee
where the Axial and Vector currents are defined as 
\be
V_\mu(x) = \bar\psi(x) \gamma_\mu \psi(x)\; , \qquad 
A_\mu(x) = \bar\psi(x) \gamma_\mu \gamma_5 \psi(x) = -i  \epsilon_{\mu\nu}
V_{\nu}\; ,
\ee 
and the scalar and pseudoscalar densities are
\be
S(x) = \bar\psi(x) \psi(x) \; , \qquad\qquad 
P(x) = \bar\psi(x) \gamma_5 \psi(x)\; .
\ee
It is a peculiar property of the two-dimensional space-time that the vector 
and the axial currents are not independent of each other.

In the massless case a free fermion field 
factorizes and the Schwinger photon acquires a mass due to the
anomaly. The photon's field $\Phi$ is 
\be
V_\mu = \frac{1}{\sqrt{\pi}}\epsilon_{\mu\nu}\partial_\nu\Phi
\ee
and the vector current correlator 
\be
\langle V_\mu(x) V_\nu(y)\rangle_0
\ee
is the same as for a free massive propagator with mass $(\mu_0/g)^2 = 1/\pi$. 
The chiral condensate is 
\be
\frac{\langle \bar \psi \psi \rangle_0}{g} = -\frac{e^{\gamma}}{2\pi^{3/2}}  
\ee
where $\gamma=0.577216\dots$ is Euler's constant. Note that the formation of 
the fermion condensate, as in one-flavor QCD, does not imply
spontaneous symmetry breaking; the $U(1)$ chiral symmetry is already
broken by the anomaly. 

For small masses the corrections to the massless 
results can be obtained 
using chiral perturbation theory \cite{coleman,adam} and the Schwinger mass
at the first order is given by 
\be
\Big(\frac{\mu_1}{g}\Big)^2 = \frac{1}{\pi} - 4 \pi
\frac{\langle \bar \psi \psi \rangle_0}{g} \Big(\frac{m}{g}\Big)  
\label{eq:massonef}
\ee

\subsection{The Schwinger model with $N_f=2$.}
\label{sec:schw_model_sub2}

In the model with two degenerate massive flavors,
the classical theory has a $U(1)_V \times SU(2)_V$
vector symmetry, an axial $SU(2)_A$ symmetry softly broken 
by the mass term and a $U(1)_A$ symmetry broken by the quantum corrections.
The Ward identities corresponding to the axial and vector $U(1)$
symmetries are analogous to the previous case. The 
Ward identities associated to the non-singlet axial and vector 
currents are given by
\be
\partial_\mu V^a_\mu = 0  
\ee
and 
\be\label{eq:AWI}
\partial_\mu A^a_\mu =  2 m P^a 
\ee
where
\be
\qquad V^a_\mu = 
\bar\psi\frac{\lambda^a}{2}\gamma_\mu \psi\; , \qquad 
A^a_\mu = \bar\psi\frac{\lambda^a}{2} \gamma_\mu \gamma_5 \psi\; , 
\qquad P^a = \bar\psi\frac{\lambda^a}{2} \gamma_5 \psi
\ee
and $\lambda^{a}\equiv\sigma^a$ are the generators of the 
$SU(2)$ in flavor space. The non-singlet axial Ward identity
(\ref{eq:AWI}) will be one of the key ingredients to test 
the properties of the overlap regularization.

In the massless limit, there are one massive singlet 
particle ($\eta$) with mass  
$(\mu_\eta/g)^2 = N_f/\pi$ and a triplet of massless particles ($\pi$).
Analogously to the $N_f=1$ model, the correlation functions 
of the vector currents are the same as for free particles.
Unlike QCD, the chiral condensate $\langle \bar\psi \psi \rangle_0=0$. 
In fact a non-zero fermion condensate would break spontaneously 
the $SU_A(2)$ chiral symmetry  of the model. But spontaneous symmetry 
breaking is not possible in two dimensions \cite{coleman2}.
In this case also, the corrections to the massless 
results can be obtained by a semi-classical analysis  
\cite{ghl} and the masses for triplet and singlet 
are\footnote{Recently a more precise computation in the limit of 
large coupling $g$ and small mass $m$ has been performed in \cite{smilga}.
It gives $c=2.008\dots$}
\ba
\label{eq:masstwof}
\frac{M_\pi}{g} & = & c \left(\frac{m}{g}\right)^{2/3} \qquad 
c = e^{2\gamma/3} \frac{2^{5/6}}{\pi^{1/6}}=2.163\dots\\
\left(\frac{M_\eta}{g}\right)^2 & = & \frac{2}{\pi} + 
\left(\frac{M_\pi}{g}\right)^2\nonumber
\ea 

\subsection{The Dirac Operator in the Overlap Formulation.}
\label{sec:schw_model_sub3}
In the lattice regularization of the Schwinger model, the fermionic
fields are defined over the sites of a square lattice, with
lattice spacing $a$, and the gauge potentials are replaced with
$U(1)$ link variables $U_\mu(x)$ defined over the oriented links
of the lattice. The Euclidean lattice action is given by 
\be   
S_L = \beta \sum_{x,\mu<\nu}\left[ 1- \mbox{Re}\, U_{\mu\nu}(x)\right] +
a^2 \sum_{i=1}^{N_f}\sum_{x,y} \bar{\psi}_i(x)\left[ \, 
(1-\frac{am}{2}) D (x,y) + m \right]\, \psi_i(y)
\label{eq:sg}   
\ee   
where $\beta = 1/(ag)^2$, $g$ being the bare coupling constant,
\be
U_{\mu\nu}(x)=U_\mu(x)U_\nu(x+a\hat\mu)U^{\dag}_\mu(x+a\hat\nu)
U^{\dag}_\nu(x)
\label{eq:umn}   
\ee
$\hat\mu$ being a unit vector in
direction $\mu$, 
and $D$ is the lattice Dirac operator in the overlap formulation,
as introduced by Neuberger.  Occasionally we will also refer to $D$
as the Neuberger-Dirac operator.  $D$ is constructed as follows.
One starts from the massless Wilson-Dirac operator
\be\label{eq:WDO}
D_W = \frac{1}{2} \gamma_\mu (\nabla_\mu + \nabla^*_\mu)
-\frac{1}{2} a \nabla^*_\mu\nabla_\mu
\ee
where $\nabla_\mu$ and $\nabla^*_\mu$ are the forward and backward lattice
derivative, i.e.
\ba
\nabla_\mu \psi(x) & = & \frac{1}{a}[U_\mu(x) \psi(x + a\hat\mu) -
\psi(x)]\nonumber\\
\nabla_\mu^* \psi(x) & = & 
\frac{1}{a}\left[ \psi(x) - U^{\dagger}_\mu(x-a\hat\mu) 
\psi(x-a\hat\mu) \right]
\label{eq:derivative}
\ea
One performs a polar decomposition of $D_w-1/a$, expressing this
operator in terms of a unitary operator $V$ and its modulus
\be\label{eq:opneub}
D_W -\frac{1}{a} = V \left[\left( D_W^\dagger -\frac{1}{a}\right) 
\left( D_W -\frac{1}{a} \right) \right]^\frac{1}{2}
\ee
 From Eq.~\ref{eq:opneub} it follows that $V$ is given by
\be\label{eq:v}
V= \left(D_W -\frac{1}{a}\right) \left[\left( D_W^\dagger -\frac{1}{a}\right) 
\left( D_W -\frac{1}{a} \right) \right]^{-\frac{1}{2}}
\ee
Finally, the Neuberger-Dirac operator for a fermion of bare mass $m$
is given by
\be\label{eq:opneub2}
D= \left(\frac{1}{a}-\frac{m}{2}\right)\left( 1 + V \right)+m
\ee
 
The Neuberger-Dirac operator 
satisfies the $\gamma_5$-Hermiticity condition
\be\label{eq:gamma5H}
D^{\dagger} = \gamma_5 D \gamma_5
\ee
Moreover, for $m=0$ it satisfies the Ginsparg-Wilson (GW) relation \cite{GW} 
\be\label{eq:GW}
\gamma_5 D^{-1} + D^{-1} \gamma_5 = a \gamma_5 
\ee
which implies that the fermion action (\ref{eq:sg}) at finite lattice 
spacing is invariant under the following continuous
symmetry~\cite{luscher}
\be\label{eq:luscher}
\delta \psi  = \gamma_5 (1- a D)\psi, \qquad 
\delta \bar\psi  = \bar \psi \gamma_5   
\ee
This can be interpreted as the lattice form of chiral symmetry.
The corresponding flavor non-singlet chiral transformations are defined
including a flavor group generator in Eq.~(\ref{eq:luscher}). 

The $\gamma_5$-Hermiticity condition (\ref{eq:gamma5H})
and the GW relation (\ref{eq:GW}) imply further algebraic relations 
which can turn out to be useful in numerical simulations \cite{Nied}. 
In particular
\be\label{eq:beautiful}
D + D^\dagger  = 
a D D^\dagger  = a D^\dagger D 
\ee
hence $D$ and $D^\dagger$ commute, i.e.~$D$ is normal and therefore 
the eigenvector are mutually orthonormal.

It is important to stress that locality, the absence of 
doubler modes and the correct classical continuum limit are not 
guaranteed by the GWR in Eq.~(\ref{eq:GW}). Indeed, there exist 
lattice fermion actions which satisfy the GWR but which
do not meet the above requirements \cite{chiu}. 
The Neuberger operator 
satisfies all the above requirements and is local in the weak coupling regime,
as shown in Ref.~\cite{pilar}.

The geometrical definition of the topological charge on the lattice is
\be
Q = \frac{1}{2\pi} \mbox{Im} \sum_{x}  \log(U_{12})
\ee
The flavor-singlet chiral transformations in 
Eq.~(\ref{eq:luscher}) leads to chiral Ward identities analogous to the 
ones in Eq.~(\ref{eq:WIs}) and the anomaly term arises from the 
non-invariance of the fermion integral measure \cite{luscher}.
The non-singlet chiral transformations analogous to Eq.~(\ref{eq:luscher})
and the locality of the Neuberger-Dirac operator imply 
axial Ward identities analogous to Eq.~(\ref{eq:AWI}) 
and therefore the quark mass
renormalizes only multiplicatively, i.e.~the critical bare mass is zero,
the corresponding conserved axial and vector currents do not need
renormalization and there is no mixing between 4-fermion operators in
different chiral representations~\cite{hasenfratz2,cg}.
Finally, the chiral condensate requires a subtraction
and is defined as 
\be\label{eq:condensate}
\chi = 
- \langle \bar \psi  (1-\frac{a}{2} D)\psi \rangle
\ee

\section{Numerical Results from an Exact Calculation}
\label{sec:exact}
In order to test our method of approximation and, at the same time,
to increase the body of information on the lattice Schwinger model,
we performed an extensive simulation with the exact calculation of 
Neuberger-Dirac operator and its inverse.  
For previous numerical work on the Schwinger model in the
overlap regularization see~\cite{nhv}-\cite{bieta}.  We considered 
the Schwinger model with $\beta=6$ on a square lattice with
$N_x = N_y =24$.  On a lattice of this size, the discretized Dirac 
operator is a complex matrix of dimension $1152 \times 1152$, for which
we could use full matrix algebra subroutines without excessive
burden on the resources available to us (the whole calculation
used approximately 20,000 processor hours on the Boston
University SGI/Cray Origin 2000 supercomputer).  Having settled
on the lattice size, we selected $\beta=6$ on the basis of previous
results which indicated that, for most of the fermion masses we
were planning to consider, the lattice would span at least
a few correlation lengths.  We generated 500 configurations
of the gauge variables $U_{\mu}(x)$ distributed according
to the pure gauge measure
\be
S_G = \beta \sum_{x,\mu<\nu}\left[ 1- \mbox{Re}\, U_{\mu\nu}(x)\right]
\label{eq:pg}   
\ee
(see also Eq.~\ref{eq:umn}).  These were obtained by a standard
Metropolis Monte Carlo simulation, with 10000 upgrades of the
whole lattice between subsequent configurations.  With the
pure gauge measure the plaquettes are essentially uncorrelated, 
apart from the constraint coming from the periodic boundary conditions,
so the procedure we followed should be amply adequate to produce
independent configurations.  For calculations with one and two
flavors of dynamical fermions, we incorporated the determinant
of the lattice Dirac operator in the averages giving the observables.
While with large variations of the determinant this way of proceeding
would lead to an unacceptable variance, in the present
calculation we found the range of values taken by ${\rm Det}(D)$
to be sufficiently limited to warrant our averaging procedure
(see Fig.~\ref{fig:det} in Section~\ref{sec:determinant}.)
This is of course due to the rather small size of our system.
With larger systems one should incorporate the determinant 
(or a suitable approximation to it) in the measure used for 
the simulation.
\begin{figure}[!ht]
\begin{center}
\includegraphics[height=8cm,width=10cm]{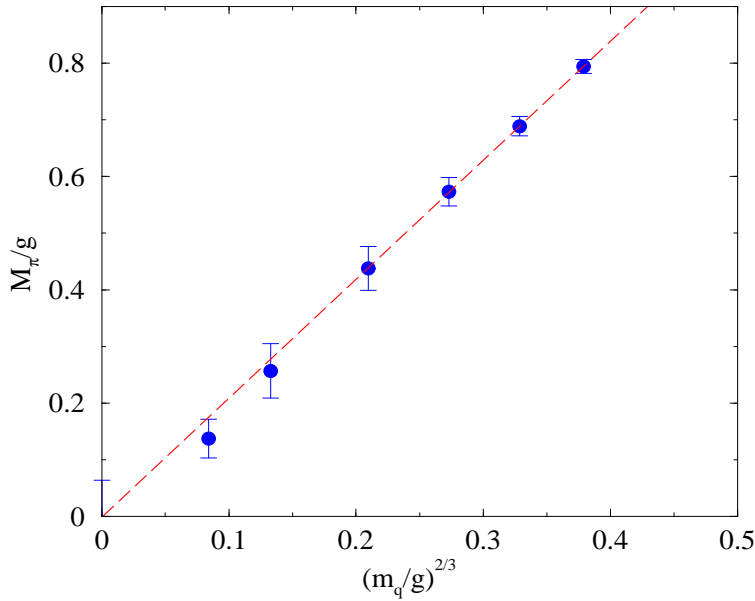}
\caption{\label{fig:mpiNf2}
\it{$M_\pi/g$ vs.~$(m/g)^{2/3}$ for the full operator for 
$N_f=2$. The dashed line represents the fit as in Eq.~\ref{eq:lastfinal} of
the four highest masses.}}
\end{center}
\end{figure} 

For each configuration, we performed a singular value decomposition
\be
D_W-\frac{1}{a}=U \Lambda \tilde U
\label{eq:svd}
\ee
where $U, \tilde U$ are unitary matrices and $\Lambda$ is diagonal,
real and non-negative.  The operator $V$ in Eq.~\ref{eq:opneub}
is given by
\be
V =U \tilde U
\label{eq:opv}
\ee
We proceeded then to the diagonalization of $V$, calculating
all its eigenvalues and eigenvectors.  From these, it is clearly
straightforward to calculate both the determinant of the Neuberger-Dirac
operator $D$ as well as its associated propagator $D^{-1}$ for any value 
of the fermion mass $m$.  The full diagonalization of $V$ is computationally
more demanding then the direct calculation of $D^{-1}$, which typically
gives also ${\rm Det} (D)$ as a by-product, but we were interested
in the actual spectrum of $V$ for comparison with the approximations
that will be discussed later.  From the fermion propagators we
calculated the meson propagators projected over zero momentum.
We focused on the vector correlators
\be
\langle \sum_{x,y,y'} \bar \psi(x,y) \gamma_2 \psi(x,y)
\bar \psi(x+t,y') \gamma_2 \psi(x+t,y') \rangle
\label{eq:mcorr}
\ee
because they are saturated by a single particle contribution in
the massless limit.  Practitioners of lattice calculations will 
certainly appreciate the value of being able to sum over all
source locations, as opposed to having to deal with selected
columns of the meson propagators. Of course we added to the 
averages the correlators obtained from the interchange of $x$
and $y$ in Eq.~\ref{eq:mcorr} for a further gain in statistics.

\begin{figure}[!ht]
\begin{center}
\includegraphics[height=8cm,width=10cm]{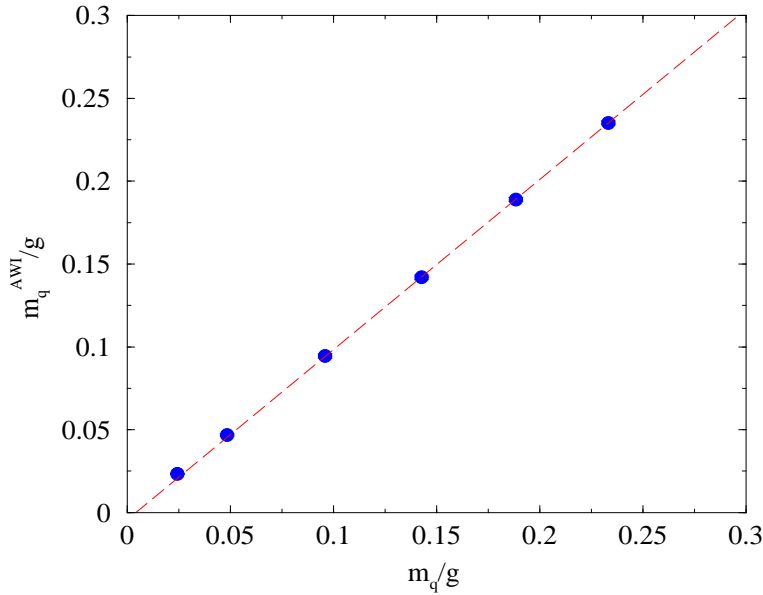}
\caption{\label{fig:rhoNf2}
\it{$m^{AWI}/g$ vs.~$m/g$ for the full operator for $N_f=2$.}}
\end{center}
\end{figure} 

\begin{figure}[!ht]
\begin{center}
\includegraphics[height=8cm,width=10cm]{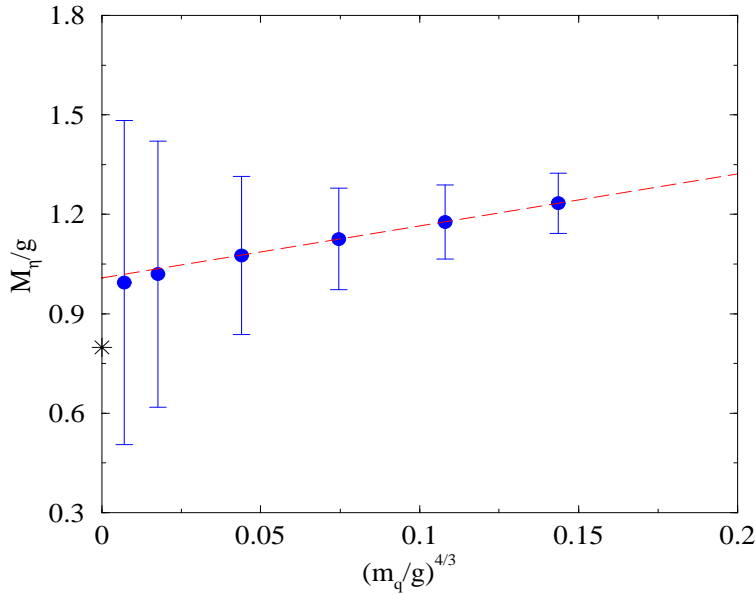}
\caption{\label{fig:etaNf2}
\it{$M_\eta/g$ vs.~$(m/g)^{4/3}$ for the full operator for $N_f=2$.}}
\end{center}
\end{figure}

\begin{figure}[!ht]
\begin{center}
\includegraphics[height=8cm,width=10cm]{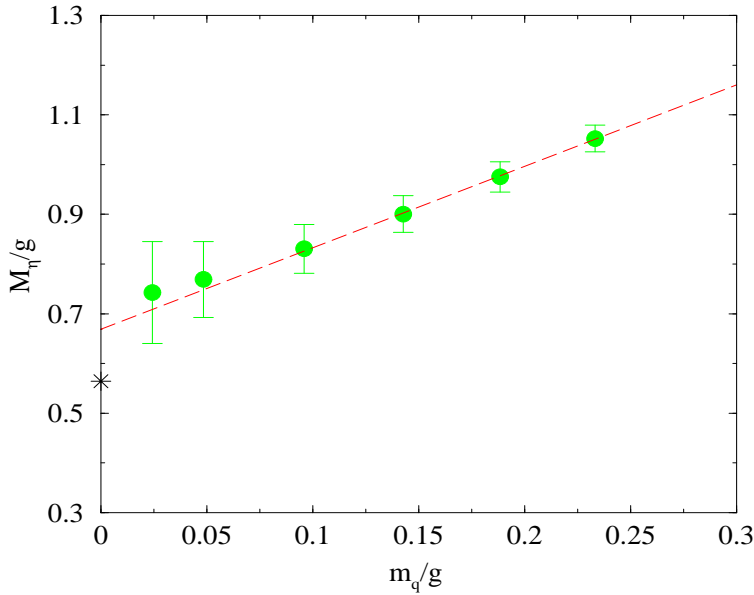}
\caption{\label{fig:etaNf1}
\it{$M_\eta/g$ vs.~$(m/g)$ for the full operator for $N_f=1$.}}
\end{center}
\end{figure}

From fits to the meson correlators we extracted the meson
masses in a standard manner.
Figures~\ref{fig:mpiNf2}, \ref{fig:rhoNf2}, \ref{fig:etaNf2},
\ref{fig:etaNf1} reproduce our results for the meson masses
as functions of the fermion mass.  The values we obtained 
for the masses are also reported in the tables included
in Sect.~\ref{sec:approx}.  The errors have been evaluated
with the jackknife method.  Figure~\ref{fig:mpiNf2}
illustrates the behavior of the isotriplet mass $M_\pi$ in the model
with two flavors.  This mass is expected to vanish for $m=0$
and chiral perturbation theory predicts an $m^{2/3}$ dependence
on $m$ (see Eq.~\ref{eq:masstwof}). The dashed line in the figure
shows the results for the fit
\be\label{eq:lastfinal}
M_\pi/g = A + B (m/g)^{2/3}
\ee
for the four heaviest masses,
which satisfy the condition $N_x M_\pi>4$ and therefore are less
likely to be affected by finite volume effects. This  
fit gives $A=-0.001(65)$ and $B=2.10(14)$.
We also performed the fit
\be\label{eq:afterfinal}
M_\pi/g = C (m/g)^{\gamma}
\ee
which gives $C=2.10(17)$ and $\gamma=0.67(7)$. These results confirm in 
an impressive manner the chiral properties of the Neuberger operator  
and the mass dependence expected by analytical calculations \cite{ghl,smilga}.

Equally gratifying
is the comparison of the value of the fermion mass $m$ in the Lagrangian
with the value $m^{AWI}$ that can be extracted, up to discretization effects, 
from the lattice analog of 
the axial Ward identity in Eq.~\ref{eq:AWI}. As shown in Fig.~\ref{fig:rhoNf2},
the intercept and the slope of the linear fit 
through the four heaviest masses are compatible with zero and
one respectively.

In Fig.~\ref{fig:etaNf2} we illustrate the behavior of the singlet
mass $M_\eta$ versus $m^{4/3}$, always for $N_f=2$.  The singlet meson 
propagator is given by the difference of the connected and disconnected 
terms in the correlator and, because of the cancellations, the errors 
are larger than in the triplet case.  The numerical results are, however, 
consistent with the theoretical prediction of Eq.~\ref{eq:masstwof}.

Finally, in Fig.~\ref{fig:etaNf1} we display the singlet mass in the
single flavor case.  The errors are smaller than in the two flavor
case because of the different relative weight of the connected and 
disconnected contributions.  Again we find reasonable
agreement between numerical results and the theoretical
prediction.

\section{Approximation of the Neuberger-Dirac Operator}
\label{sec:approx}
Anybody who has ever seen a plot of the spectrum of eigenvalues
of the Wilson-Dirac operator cannot but be left with the
impression that there is a huge redundancy of states.
The projection over a translated unitary circle done
by the Neuberger-Dirac operator avoids the problem of
mode doubling, but preserves the overall count of states.
Yet, it would appear that the physical properties of
the system should be determined by the eigenstates of $D$ with
eigenvalues in the neighborhood of $\lambda=m$ (i.e.~the eigenstates 
of $V$ with eigenvalue $\lambda_V$ closest to $-1$, 
cfr.~Eqs.~\ref{eq:v}, \ref{eq:opneub2}), 
since these are the states with a smooth long
range behavior, expected to go over the physical states of
the continuum in the limit $a \to 0$.  This suggests
that it should be possible to reconstruct 
the physical observables from these states alone.
The way in which such states contribute to physical observables
has been studied in the literature~\cite{lowmodes}.
Here we would like to make the point that the overlap formulation
is particularly well suited for the implementation
of the above approximation, since the unitarity of $V$
provides, in some sense, a checkpoint on the approximation itself.
Of course, one must be careful in attempting any approximation
based on neglecting the short-wavelength part of the spectrum,
even if the corresponding states are largely lattice artifacts,
since one knows that in quantum field theory the infrared
and ultraviolet components of the spectrum are subtly related.
Thus, the chiral eigenstates with $\lambda_V=-1$ (``zero modes''),
which $V$ exhibits in presence of a gauge field with non-trivial topology, 
find their counterpart in states with opposite chirality at $\lambda_V=1$.
Here too, however, the special features of the overlap formulation
come to the rescue, since, as we will show, both the presence
of zero modes and the correspondence between $\lambda_V=-1$
and $\lambda_V=1$ eigenstates of opposite chirality can be
preserved by the projection over a subset
of physical states.  In this section we will illustrate 
such a projection by comparing the approximate results for the 
observables with those of the exact calculation.

\begin{figure}[!ht]
\begin{center}
\includegraphics[height=8cm,width=8cm]{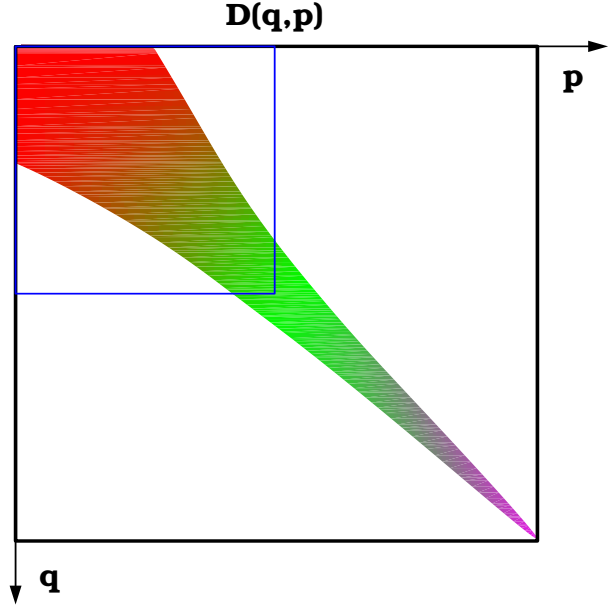}
\caption{\label{fig:2rho}
\it{Structure of the Wilson operator in momentum space and 
in a smooth gauge.}}
\end{center}
\end{figure}

One possible scheme of approximation which is computationally
very convenient consists in performing a projection over states
of low momentum in Fourier space, after gauge fixing to a
smooth gauge field configuration.  In a smooth gauge, because
of the suppression of short-wavelength fluctuations due to
asymptotic freedom, one expects the structure of the Wilson operator 
come more and more diagonal in momentum space for increasing
momenta.  This notion is schematically illustrated in Fig.~\ref{fig:2rho}
and also underlies the technique of Fourier acceleration for
the calculation of quark propagators \cite{Wilson}.  Accordingly, we
implemented the following approximation.

We fixed the gauge on all configurations to the Landau gauge 
by demanding that the function
\be
G = \sum_{x, \mu} {\rm Re}[ U_\mu(x)]  
\label{eq:lgauge}
\ee
be maximal.  A relaxation procedure produces several local maxima
(Gribov copies).  Among all these configurations, we selected the
one that produced the absolute maximum for $G$, subject to the
further constraint that for all $x$ and $\mu$ ${\rm Re} U_\mu(x) \ge 0.5$.
Some care must be exerted with the configurations that have non-vanishing
topological number $Q$.  These configurations cannot be brought to
a uniformly smooth gauge.  To discuss them further it is convenient
to introduce the notation
\ba
U_{\mu}(x)=e^{\imath \theta_{\mu}(x)} \label{eq:thetamu}\\
U_{\mu\nu}(x)=e^{\imath \theta_{\mu\nu}(x)} \label{eq:thetamunu}
\ea
Then, with $Q\ne0$ there must be one or more plaquettes where
\be
\theta_{\mu\nu}(x) = \theta_{\mu}(x)+ \theta_{\nu}(x+a\hat \mu)-
\theta_{\mu}(x+a \hat \nu)- \theta_{\nu}(x) + 2 m \pi
\label{eq:deltatheta}
\ee
with $m$ a non-zero integer. This implies that for such plaquettes
\be
{\rm Abs}[\theta_{\mu}(x)+ \theta_{\nu}(x+a\hat \mu)-
\theta_{\mu}(x+a \hat \nu)- \theta_{\nu}(x)] \ge \pi
\label{eq:abstheta}
\ee
and thus the corresponding gauge variables cannot be all close to unity.
We will call these plaquettes the ``hot spots'' of the gauge configuration.
Their location is gauge dependent but, as long as $Q\ne0$, they cannot be
eliminated.  For the configurations $C_Q$ with $Q\ne0$ we fixed the gauge as
follows.  First we created a configuration $C'$ with the same value of $Q$
by superimposing gauge field configurations in the Landau gauge with 
uniform $U_{\mu\nu}(x)$ and $Q=1$.  We placed the hot spots of this
configurations in the locations where the action of the original
field configuration was smaller, using a weighting procedure that
minimized the overlap of hot spots.  Then we brought to the Landau gauge 
the gauge field configuration $C''$ obtained subtracting $C'$ from $C_Q$.
Finally we superimposed $C'$ to the gauge transformed $C''$. The resulting
configuration is gauge equivalent to the original $C_Q$, is 
approximatly in the
Landau gauge and is ``reasonably smooth'', the latter statement being
justified a posteriori by the application of our approximation.
Clearly the procedure described above relies heavily on the Abelian 
nature of the gauge field and will have to be generalized to handle
non-Abelian systems.

After gauge fixing, we project the lattice Wilson-Dirac operator over 
the subspace $\cal F$ of Fourier space spanned by the eigenvectors
with lowest momenta.  For the results we present in this paper,
the projection was done over the subspace defined by
\be
(p_x a)^2+(p_y a)^2 \le (p_c a)^2
\label{eq:subspace}
\ee
where the cut-off momentum $p_c$ was chosen so that the total number
of momenta included in the projection equals 145, i.e.~a fraction
$z \approx 1/4$ of the total number of momenta. (We could not choose $p_c$ 
in such a way as to project over exactly one fourth of the space, 
since the number of momenta satisfying Eq.~\ref{eq:subspace}
is always odd.)  Accordingly, the dimensionality of $\cal F$
is 290, out of a total dimensionality of the fermion field equal 
to 1152 ($24 \times 24$ lattice sites $\times 2$ spin degrees of freedom).  
This corresponds to an effective reduction of a factor of two for each 
dimension of the lattice.  As a matter of fact, we experimented with even
smaller values of $z$ and found that, with $Q=0$, $z$ can
be chosen substantially smaller than $1/4$ without spoiling the
good quality of the approximation. For configurations with a non-trivial
topology the infrared properties of the Wilson operator are still 
well reproduced on the subspace. However
the real doubler modes of the truncated Wilson operator are 
shifted towards the infrared region and, as they pass the projection 
point, the association between topology and chirality is
lost. 

The approximation to the Neuberger-Dirac operator is obtained
following the construction of Eqs.~\ref{eq:opneub} and \ref{eq:v},
where $D_W$ and $V$ are replaced by their projections over
$\cal F$: $\tilde D_W$ and $\tilde V$ (see also \cite{GHRlat}).  
If we denote by $D_W^{(0)}$ the projection of the free Dirac operator
over the complement of $\cal F$ and by $V^{(0)}$ the corresponding
unitary factor in its polar expansion, we finally take
\be
V_{approx}= \tilde V \oplus V^{(0)}
\label{eq:vapprox}
\ee
as our approximation to $V$.  More specifically, given any fermionic 
vector $\psi$, we project it over $\cal F$ and its complement: 
$\psi=\psi_F+\psi_{\bar F}$. $V_{approx} \psi$ is then given by 
$\tilde V \psi_F + V^{(0)}_{\bar F}$.The approximate form of the
Neuberger-Dirac operator follows from Eq.~\ref{eq:opneub2}.
\begin{figure}[!ht]
\begin{center}
\includegraphics[height=8cm,width=11cm]{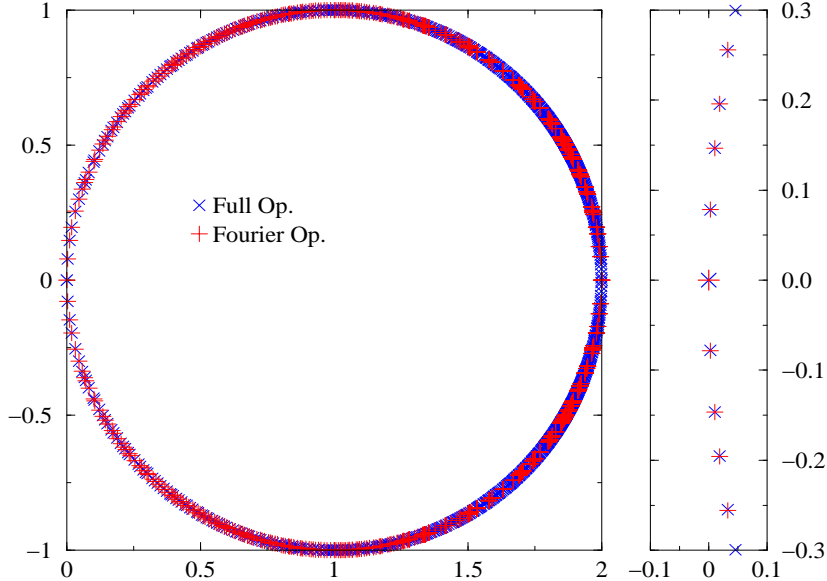}
\caption{\label{fig:spectrum}
\it{Comparison of spectra on a configuration with $Q=1$.}}
\end{center}
\end{figure} 

The first check on the approximation is that it should reproduce
well the spectrum of long range modes, i.e.~those with 
$V$ eigenvalues  in the neighborhood of -1.  In particular,
configurations with $Q\ne 0$ should have $|Q|$
eigenvectors with $V$ eigenvalue exactly equal to -1.
This turned out to be the case.  With the chosen cut-off $p_c$
(See Eq.~\ref{eq:subspace}) the approximation consistently gives  
satisfactory results for the spectrum of long range modes. 
We display in Fig.~\ref{fig:spectrum}
the spectrum of eigenvalues of the exact $1+V$ (x symbols - blue)
and the approximate $1+ \tilde V$ (plus symbols - red) for a configuration
with $Q=1$.   The matching in the physical part of the spectrum, 
close to $0$, is excellent (see the blow-up at the right of the figure).
Also, as we anticipated above, the approximation preserves
the presence of a zero mode, which is a chiral eigenstate.
It is worthwhile to notice that $\tilde V$ is constructed from a
projected Wilson operator $\tilde D_W$ which has the same $\gamma_5$
transformation properties as the original Wilson operator.  Thus
$\tilde V$ will have the same $\gamma_5$ transformation properties as
$V$ and, in particular, its eigenvectors with eigenvalue $-1$ will be
matched by a corresponding number of eigenvectors with eigenvalue $1$.
Since $V^{(0)}$ has no eigenvectors with eigenvalues $\pm 1$, this
carries over to the $V_{approx}$.

\begin{figure}[!ht]
\begin{center}
\includegraphics[height=8cm,width=11cm]{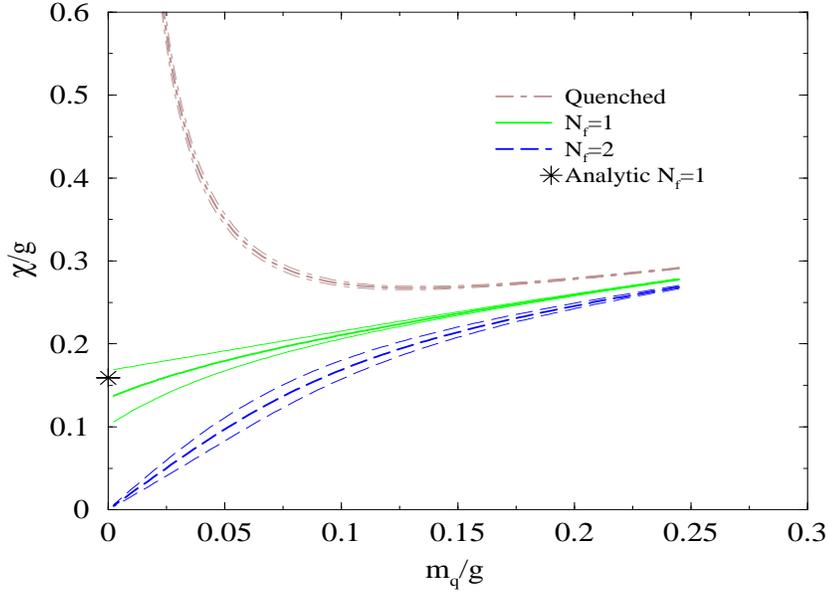}
\caption{\label{fig:chcondfull}
\it{Chiral condensate from the exact calculation.}}
\end{center}
\end{figure} 
\begin{figure}[!ht]
\begin{center}
\includegraphics[height=8cm,width=11cm]{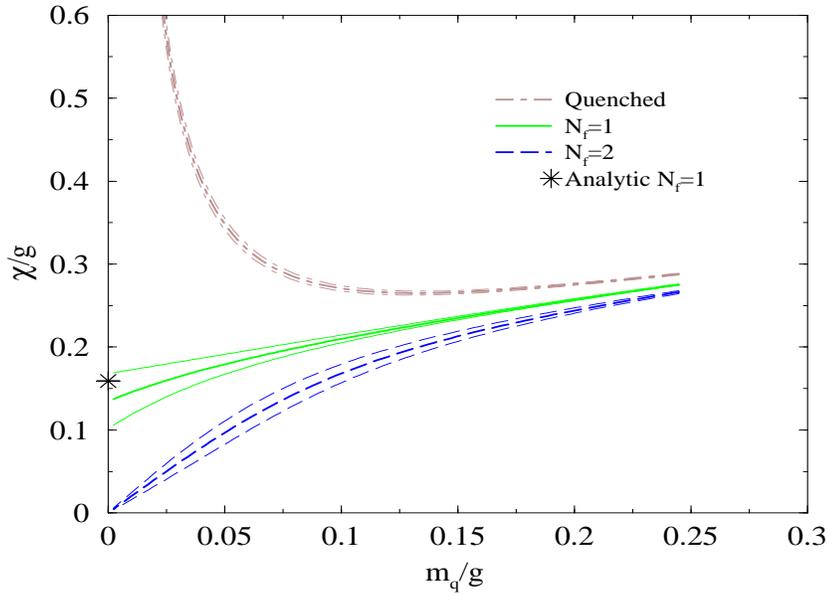}
\caption{\label{fig:chcondfourier}
\it{Chiral condensate with the Fourier approximation.}}
\end{center}
\end{figure} 

Closely related to the spectrum is the value of the chiral
condensate.  It is instructive to reexpress the condensate of
Eq.~\ref{eq:condensate} in terms of the eigenvalues of $V$.  
With $a=1$ we find
\ba
\chi = - \frac{1}{N_f}
\langle \bar \psi (1-\frac{1}{2} D) \psi\rangle = 
\frac{1}{N_x N_y}\frac{1}{1 - m/2}\frac{\langle  {\rm Det}(D)^{N_f} {\rm Tr} \Big[D^{-1}-
\frac{1}{2}\Big] \rangle_U}{\langle {\rm Det}(D)^{N_f} \rangle_U}
\label{eq:trace}
\ea
(We use the subscript $U$ in the r.h.s.~of this equation as a reminder
that the averages are quenched averages over the gauge field.  See also
the discussion following Eq.~\ref{eq:pg}.)
Using Eq.~\ref{eq:opneub2} we get
\be
\chi = \frac{1}{N_x N_y}\frac{\langle {\rm Det}(D)^{N_f} \Big[\frac{|Q|}{m}  
+\frac{m}{2} \sum_j \frac{1-\cos \phi_j}{(1+m^2/4)+(1-m^2/4)\cos \phi_j} \Big]
\rangle_U}{\langle {\rm Det}(D)^{N_f} \rangle_U}
\label{eq:trace3}
\ee
where we separated in the sum the $|Q|$ pairs of eigenvalues
equal to -1 and 1 from the others, which occur in pairs of complex 
conjugate values $\exp(\pm \imath \phi_j)$.

Figure~\ref{fig:chcondfull} reproduces
the result of the exact calculation, Fig.~\ref{fig:chcondfourier}
the result obtained by inserting in Eq.~\ref{eq:trace3} the eigenvalues
obtained with the Fourier approximation.  The two sets of results
are indistinguishable (we did not plot them on the same graph, because
one set of lines would have covered the other).  The crosses on the
two figures show the analytic result for $m=0$, $N_f=1$.

\begin{figure}[!ht]
\begin{center}
\includegraphics[height=8cm,width=10cm]{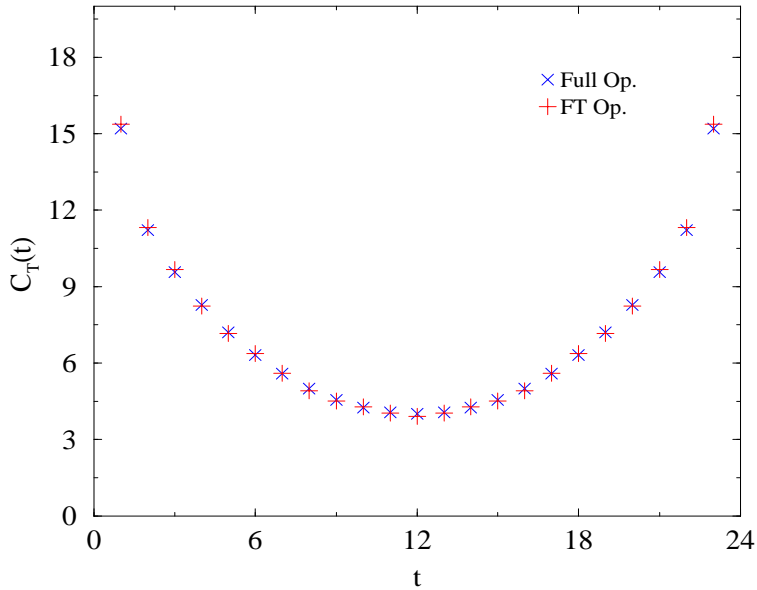}
\caption{\label{fig:schw_6.0_1006_TRIP2}
\it{Results for $\gamma_2$-triplet propagator on a configuration with $Q=0$.}}
\end{center}
\end{figure} 
\begin{figure}[!ht]
\begin{center}
\includegraphics[height=8cm,width=10cm]{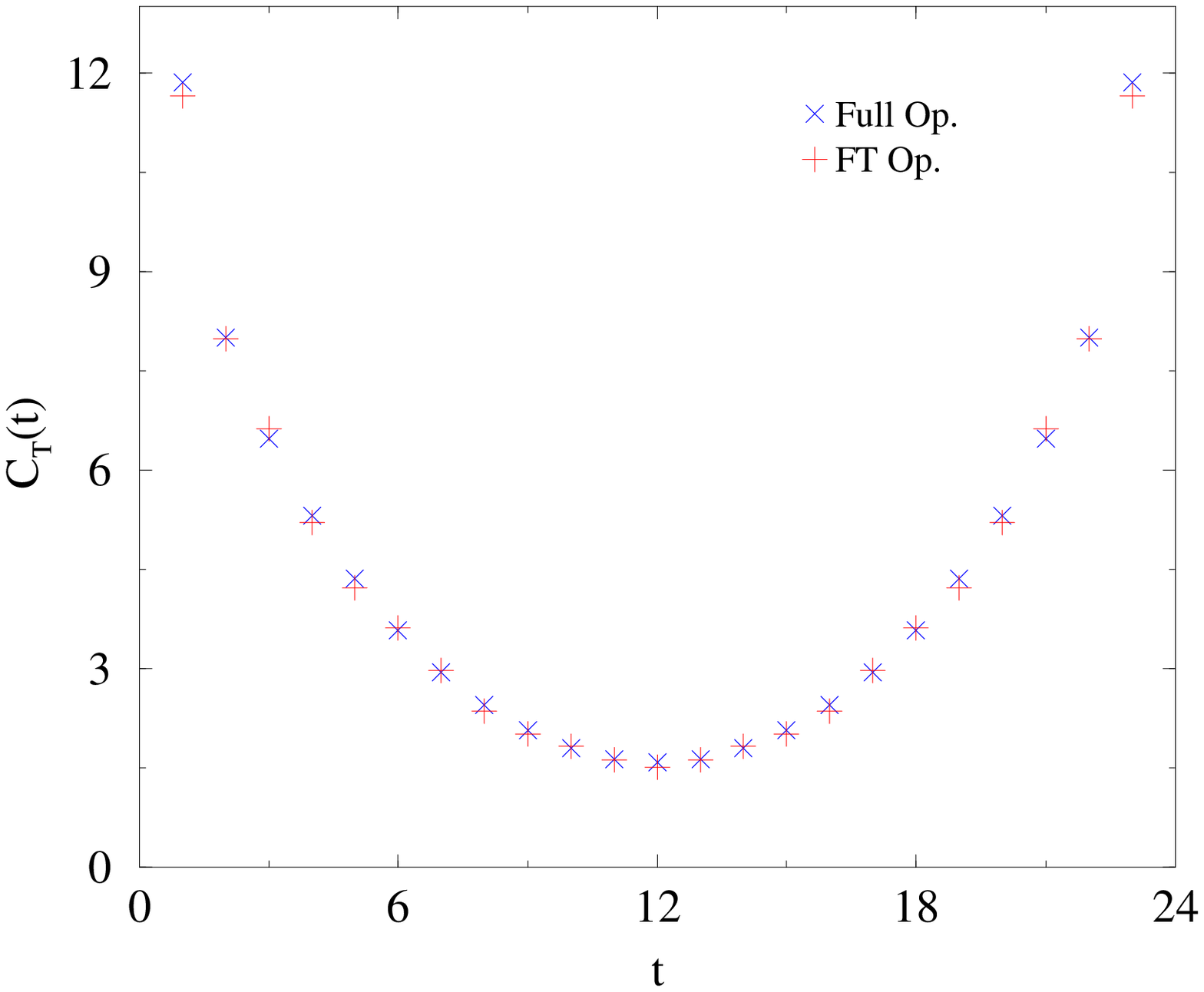}
\caption{\label{fig:schw_6.0_1003_TRIP2}
\it{Results for $\gamma_2$-triplet propagator on a configuration with $Q=-1$.}}
\end{center}
\end{figure} 
\begin{figure}[!ht]
\begin{center}
\includegraphics[height=8cm,width=10cm]{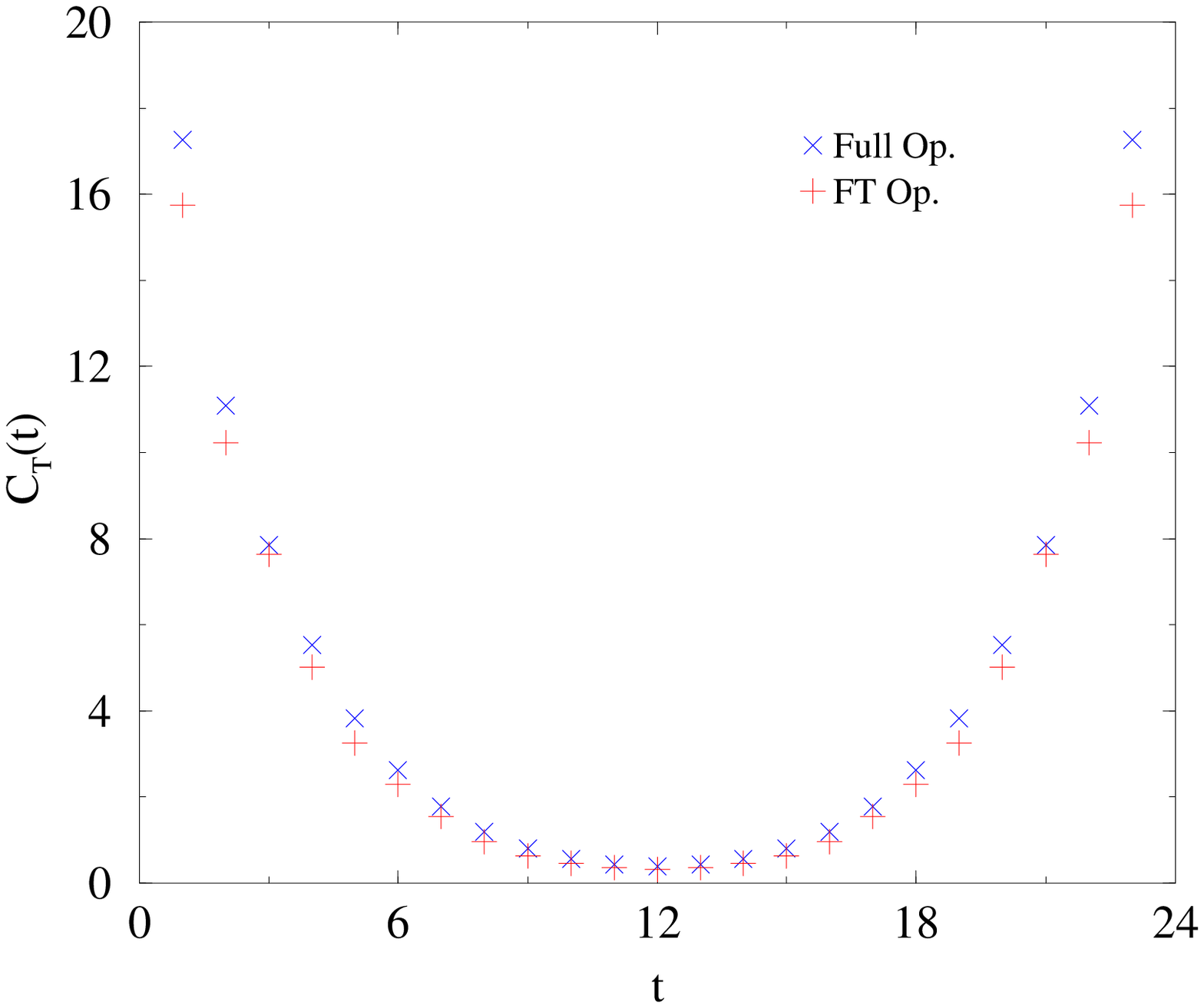}
\caption{\label{fig:schw_6.0_1019_TRIP2}
\it{Results for $\gamma_2$-triplet propagator on a configuration with $Q=4$.}}
\end{center}
\end{figure} 
\begin{figure}[!ht]
\begin{center}
\includegraphics[height=8cm,width=10cm]{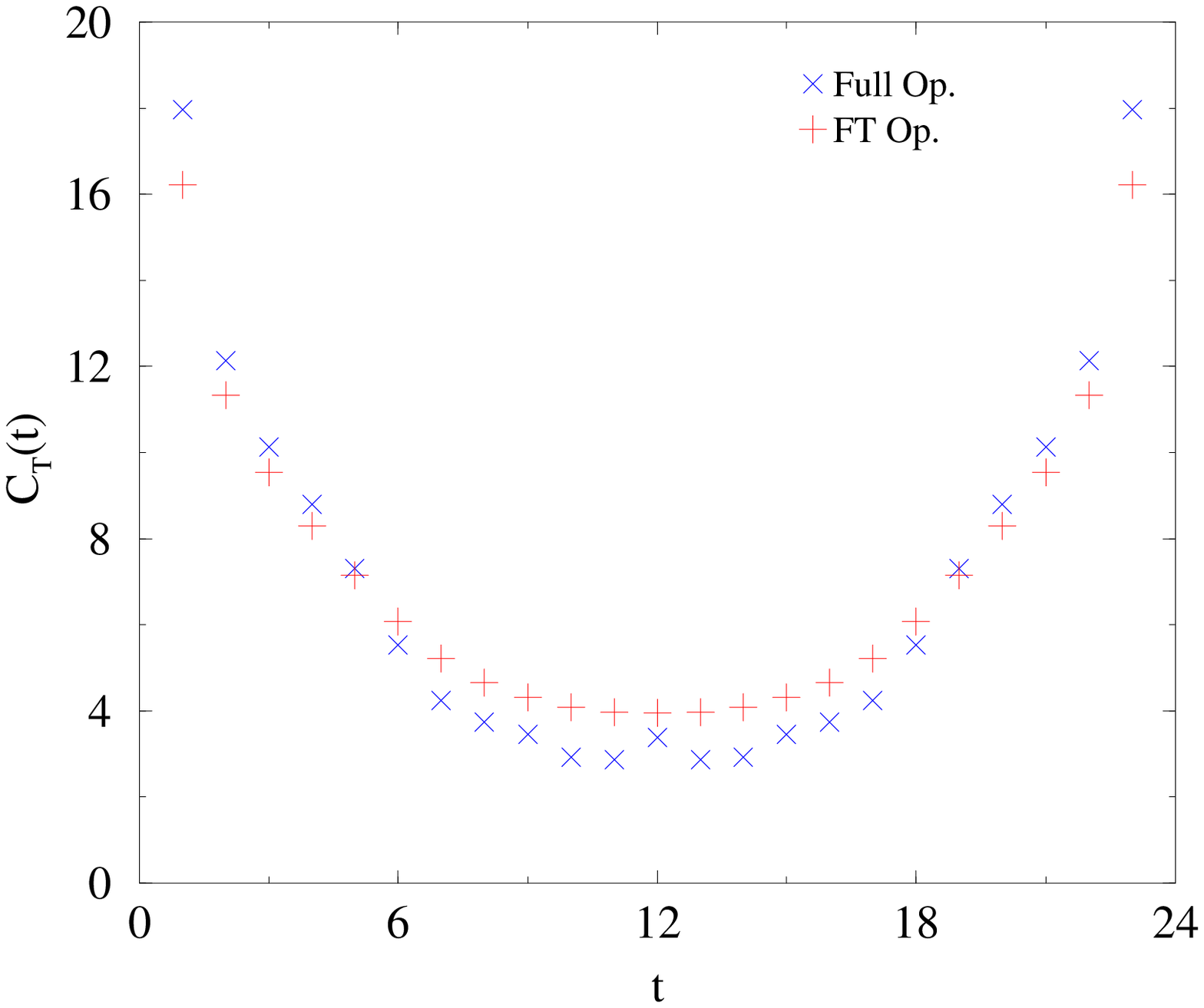}
\caption{\label{fig:schw_6.0_1006_SING2}
\it{Disconnected part of the $\gamma_2$-singlet 
propagator on a configuration with $Q=0$.}}
\end{center}
\end{figure} 
\begin{figure}[!ht]
\begin{center}
\includegraphics[height=8cm,width=10cm]{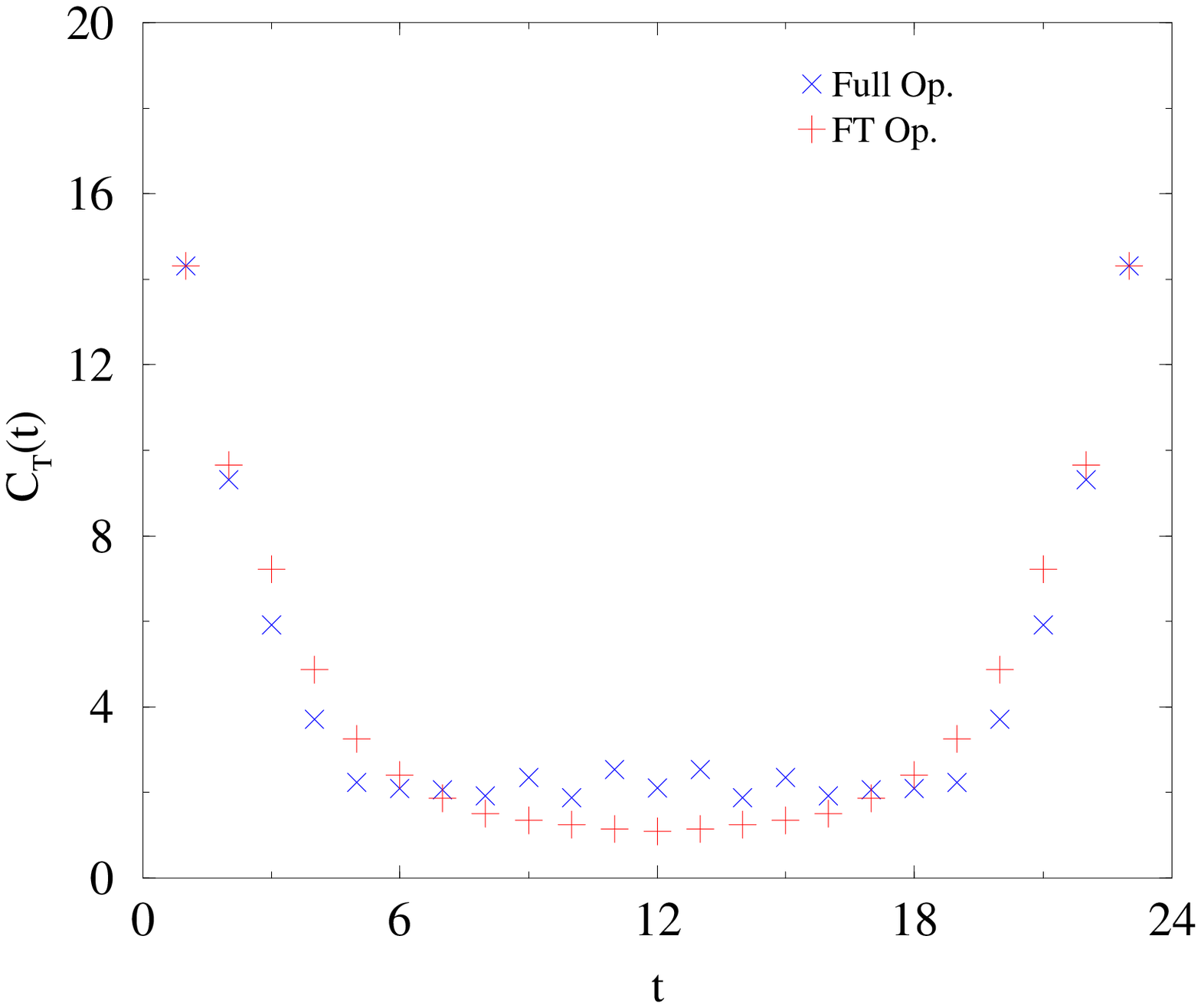}
\caption{\label{fig:schw_6.0_1003_SING2}
\it{Disconnected part of the $\gamma_2$-singlet 
propagator on a configuration with $Q=-1$.}}
\end{center}
\end{figure} 
\begin{figure}[!ht]
\begin{center}
\includegraphics[height=8cm,width=10cm]{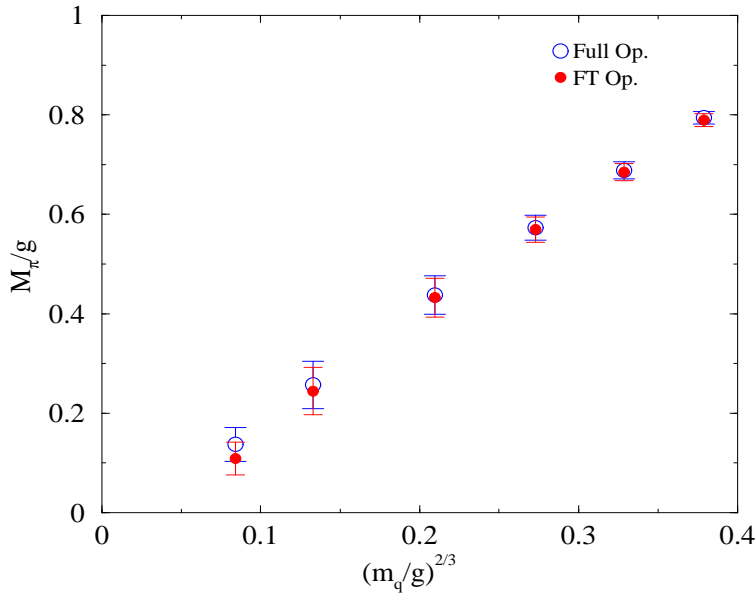}
\caption{\label{fig:mpiNf2gfix}
\it{$M_\pi/g$ vs.~$(m/g)^{2/3}$ for the full operator 
and the gauge-fixing approximation  for $N_f=2$.}}
\end{center}
\end{figure} 

Another test of the approximation is obtained by comparing 
configuration by configuration the propagators for fermion
bilinears.
Our approximation gives quite satisfactory results for the connected
Green's functions.
We reproduce in Figs.~\ref{fig:schw_6.0_1006_TRIP2},  
\ref{fig:schw_6.0_1003_TRIP2} and \ref{fig:schw_6.0_1019_TRIP2}
the propagator of the $\gamma_2$-triplet bilinear
in three randomly chosen configurations with $Q=0, -1$ and $4$, respectively.   
We see that propagators obtained from the approximation to the
Neuberger-Dirac operator compare quite well
with those obtained from the exact operator,
even with $Q=4$. 

Our results for the disconnected Green's functions are not conclusive.
We reproduce in 
Figs.~\ref{fig:schw_6.0_1006_SING2},~\ref{fig:schw_6.0_1003_SING2}
the disconnected parts of the propagator in the same configurations as used 
for Figs.~\ref{fig:schw_6.0_1006_TRIP2},~\ref{fig:schw_6.0_1003_TRIP2}.
The approximate propagators are smoother than the exact ones, which can
be understood as an effect of the truncation over the short wavelength modes,
but not incompatible in an average sense.  Of more concern is that we found
the fluctuations of the approximate disconnected propagators to be 
substantially larger than in the exact calculation.  The large
fluctuations of the disconnected components in the exact calculation already 
makes the error in the singlet masses rather big.  The even larger
fluctuations of the disconnected components in the approximate calculation 
prevented us from obtaining in this case a meaningful result for the 
singlet masses.  

We reproduce in the tables all the masses which we were able to calculate.
We see that whenever we could get a mass value from the Fourier approximation,
the result turned out in reasonably good agreement with the exact
calculation. A fit to Eqs.~\ref{eq:lastfinal} and \ref{eq:afterfinal} 
of the approximate 
calculation gives the following results 
for the parameters: $A=-0.005(65)$, $B=2.10(15)$, $C=2.10(17)$ and
$\gamma=0.67(7)$ with excellent agreement with the results of the 
full calculation, presented immediately after Eqs.~\ref{eq:lastfinal} 
and \ref{eq:afterfinal}.
A comparison between masses in the exact and approximate calculations is
also presented in Fig.~\ref{fig:mpiNf2gfix}.

\begin{table}[!ht]
\begin{center}
\begin{tabular}{||l|ll||}
\hline\hline
\multicolumn{3}{||c||}{$\beta=6.0$, $V=24^2$} \\
\hline
$m/g$  & $m_{\pi}/g$ & $m^{AWI}_q/g$  \\
\hline
\hline
\multicolumn{3}{||c||}{Full Operator} \\
\hline
       0      & 0.15(1) & -0.0026(2)  \\ 
\hline
       0.0244 & 0.393(13) & 0.0234(1)     \\ 
       0.0485 & 0.442(10) & 0.0472(2)    \\ 
       0.0960 & 0.553(8) & 0.0949(3)     \\ 
       0.1427 & 0.661(6) & 0.1420(4)     \\ 
       0.1884 & 0.761(6) & 0.1884(4)     \\ 
       0.2333 & 0.857(5) & 0.2343(5)     \\ 
\hline\hline
\multicolumn{3}{||c||}{Fourier Approximation} \\
\hline
       0.0    & 0.14(1) & -0.0049(3)  \\
\hline  
       0.0244 & 0.385(13) & 0.0208(3)      \\ 
       0.0485 & 0.435(10) & 0.0427(5)      \\ 
       0.0960 & 0.548(8)  & 0.0875(8)      \\ 
       0.1427 & 0.656(7) & 0.1321(11)      \\ 
       0.1884 & 0.758(6) & 0.1761(13)      \\ 
       0.2333 & 0.855(5) & 0.2194(16)      \\ 
\hline\hline
\end{tabular}
\caption{\label{tab:Nf=0}
Masses for $N_f=0$.}
\end{center}
\end{table}

\begin{table}[!ht]
\begin{center}
\begin{tabular}{||l|l||}
\hline\hline
\multicolumn{2}{||c||}{$\beta=6.0$, $V=24^2$} \\
\hline
$m/g$  & $m_{\eta}/g$  \\
\hline\hline
\multicolumn{2}{||c||}{Analytic} \\
\hline
0 & 0.5642 \\ 
\hline
\hline
\multicolumn{2}{||c||}{Full Operator} \\
\hline
       0      & 0.67(6)  \\ 
\hline
       0.0244 & 0.74(10)      \\ 
       0.0485 & 0.77(8)       \\ 
       0.0960 & 0.83(5)       \\ 
       0.1427 & 0.90(4)       \\ 
       0.1884 & 0.98(3)       \\ 
       0.2333 & 1.05(3)       \\ 
\hline\hline
\multicolumn{2}{||c||}{Multigrid Det.} \\
\hline
        0     & 0.68(6)  \\
\hline  
       0.0244 & 0.75(10)      \\ 
       0.0485 & 0.78(7)      \\ 
       0.0960 & 0.84(5)      \\ 
       0.1427 & 0.91(4)      \\ 
       0.1884 & 0.98(3)      \\ 
       0.2333 & 1.06(3)      \\ 
\hline\hline
\end{tabular}
\caption{\label{tab:Nf=1}
Masses for $N_f=1$.}
\end{center}
\end{table}
\begin{table}[!ht]
\begin{center}
\begin{tabular}{||l|lll||}
\hline\hline
\multicolumn{4}{||c||}{$\beta=6.0$, $V=24^2$} \\
\hline
$m/g$  & $m^{AWI}$ & $m_\pi/g$ & $m_{\eta}/g$ \\
\hline\hline
\multicolumn{4}{||c||}{Analytic} \\
\hline
0 & 0 & 0 & 0.7979    \\ 
\hline
\hline
\multicolumn{4}{||c||}{Full Operator} \\
\hline
      0.00   & -0.004(2)& -0.001(65) & 1.00(25) \\ 
\hline
      0.0244 & 0.0233(2) & 0.14(3) & 1.0(5)      \\ 
      0.0485 & 0.0468(6) & 0.26(5) & 1.0(4)      \\ 
      0.0960 & 0.0945(13)& 0.44(4) & 1.1(2)      \\ 
      0.1427 & 0.1420(14)& 0.57(3) & 1.1(2)      \\ 
      0.1884 & 0.1890(13)& 0.69(2) & 1.2(1)      \\ 
      0.2333 & 0.2353(11)& 0.79(1) & 1.23(9)     \\ 
\hline\hline

\multicolumn{4}{||c||}{Fourier Approximation} \\
\hline
      0.00   & -0.002(2) & -0.005(66)  & -  \\ 
\hline
      0.0244 & 0.018(1) & 0.11(3)      & -      \\ 
      0.0485 & 0.044(1) & 0.24(5)      & -      \\ 
      0.0960 & 0.092(1) & 0.43(4)      & -      \\ 
      0.1427 & 0.139(1) & 0.57(3)      & -      \\ 
      0.1884 & 0.184(1) & 0.68(2)      & -      \\ 
      0.2333 & 0.228(1) & 0.79(1)      & -      \\ 
\hline\hline
\multicolumn{4}{||c||}{Multigrid Determinant} \\
\hline
      0.00 &  -0.004(1) & -0.002(40)  & 1.09(25)  \\ 
\hline
      0.0244 & 0.0234(1) & 0.14(2) & 1.1(4)      \\ 
      0.0485 & 0.0469(4) & 0.26(3) & 1.1(4)      \\ 
      0.0960 & 0.0946(9) & 0.44(2) & 1.1(2)      \\ 
      0.1427 & 0.1420(10)& 0.57(2) & 1.2(2)      \\ 
      0.1884 & 0.1889(9) & 0.68(1) & 1.2(1)      \\ 
      0.2333 & 0.2351(8) & 0.79(1) & 1.26(9)     \\ 
\hline\hline
\end{tabular}
\caption{\label{tab:Nf=2}
Masses for $N_f=2$.}
\end{center}
\end{table}

In closing this section let us mention a different, gauge invariant scheme 
of approximation, which we also tried.  We built the subspace $\cal F$
used for the projection of Eq.~\ref{eq:vapprox} out of the lowest 
eigenvectors of the negative of the covariant Laplacian,
namely the operator (see Eq.~\ref{eq:derivative})
\be 
-\Delta = - \sum_\mu \nabla_\mu \nabla^*_\mu
\label{eq:covlaplacian}
\ee
The lowest eigenmodes of $-\Delta$ are typically slowly varying, whereas
the highest modes exhibit rapid variations from site to site.  Thus
the subspace $\cal F$ formed out of the lowest eigenvectors of  
$-\Delta$ is well suited to represent the physical excitations of the
system.  The construction of this subspace is gauge-invariant and,
contrary to the approximation based on the Fourier transform,
does not require any gauge fixing.  Finding a sufficiently large
number of eigenvectors of $-\Delta$ is however computationally expensive.
An even more serious problem is that this approximation does not
provide any simple way to define the projection of a ``free operator''
on the complement of $\cal F$:  the form of the free Wilson operator 
is clearly gauge dependent.  Thus, for lack of better alternatives,
we replaced $V{(0)}$ in Eq.~\ref{eq:vapprox} with the identity operator,
projecting all the remaining eigenvalues to $1$.  As a consequence of this
rather drastic projection, the propagators of the bilinears turned out
to exhibit oscillations around the exact propagators, with much less 
satisfactory agreement than we found in the Fourier approximation.
While these oscillations do not necessarily rule out the possibility 
of still obtaining good values for the masses through a suitable 
fitting procedure, the worst quality of the bilinear propagators together 
with the higher computational costs dissuaded us from pursuing this 
approximation further.
 
\section{Coarse grid Approximation to the Determinant}
\label{sec:determinant}

In the previous section we explored the possibility to truncate the
Neuberger operator to its long range components on a given gauge
configuration. In the spirit of long range approximations, there is another
possibility that can be explored, which is the coarse graining of the 
Dirac operator, as done in multigrid calculations~\cite{claudio}.  

\begin{figure}[!ht]
\begin{center}
\includegraphics[height=8cm,width=8cm]{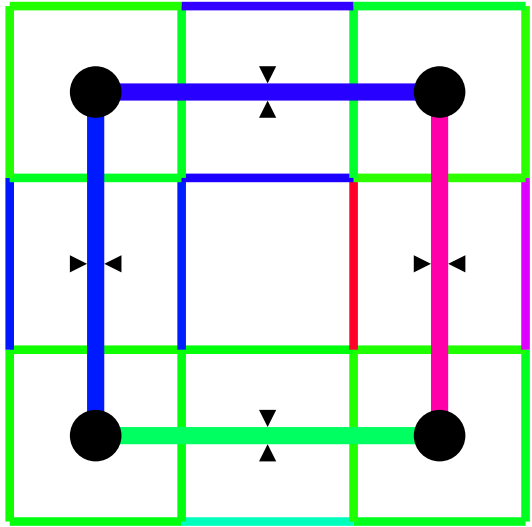}
\caption{\label{fig:block}
\it{Sketch of the blocking procedure. Original Links 
corresponding to a site in the coarse lattice are fixed
to local Landau gauge. Links
between them are averaged according to Eq.~\ref{eq:block} to give
links of the blocked lattice.}}
\end{center}
\end{figure} 

A preliminary step is the projection of the gauge field to a coarser
lattice.  Starting from an arbitrary gauge configuration on an
$N\times N$ lattice, we apply a blocking procedure to arrive at a
gauge configuration on a coarser $N/2\times N/2$ lattice designed 
in such a way that the coarser lattice should carry the long range 
features of the finer $N\times N$ lattice. The absence of additive mass
renormalization then allows one to define a Neuberger operator on the
coarser lattice, which is related to the Neuberger operator on the
finer lattice without tuning of the bare mass and reproduces all its
long range features. One could then in principle measure the Green's
functions on the coarser lattice. It would however be no trivial
problem to interpolate them back to the finer lattice compensating
for the errors introduced by coarsening.
We therefore decided only to measure the determinant on the coarse
lattice and compare it to the one on the fine lattice.

\begin{figure}[!ht]
\begin{center}
\includegraphics[height=8cm,width=10cm]{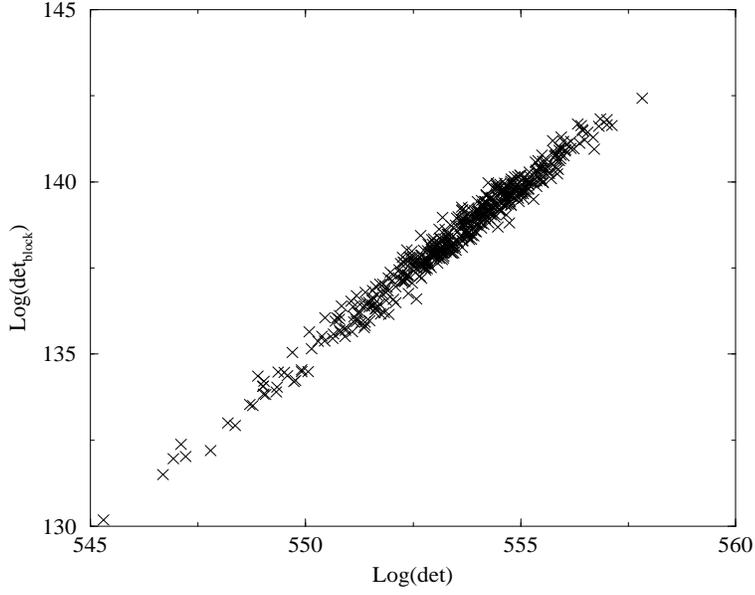}
\caption{\label{fig:det}
\it{$\log {\rm Det}(D)$ on the blocked lattice 
versus $\log {\rm Det} (D)$ on the full lattice, for lightest mass, 
$m/g=0.0244$, on the full system.}}
\end{center}
\end{figure} 

\begin{figure}[!ht]
\begin{center}
\includegraphics[height=8cm,width=10cm]{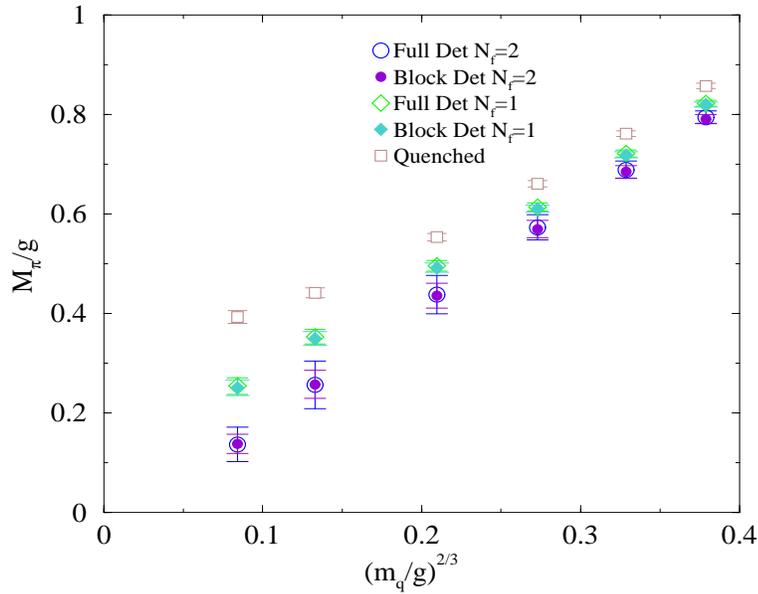}
\caption{\label{fig:mpiNf2_grid}
\it{$M_\pi/g$ vs.~$(m/g)^{2/3}$ for the full determinant
and for the blocked one.}}
\end{center}
\end{figure} 

To block a gauge field $U_{\mu}(x,t)$ on an $N\times N$ lattice down to a
gauge field $\bar{U}_{\mu}(\bar{x},\bar{t})$ on an $N/2\times
N/2$ lattice we applied the following procedure:
\begin{itemize}
\item
Divide the sites of the fine lattice into $N/2\times N/2$
blocks of $2^2$ sites. The center of each of
these blocks will correspond to a site $(\bar{x},\bar{t})$ in the coarse
lattice.
\item
Fix the gauge such that the links connecting sites within one block are as
close as possible to unity, i.e.,
\begin{equation}
\sum_{\bar{x},\bar{t}=0}^{N/2-1} \left(
U_{\hat{x}}(2\bar{x},2\bar{t})+U_{\hat{t}}(2\bar{x},2\bar{t})+
U_{\hat{x}}(2\bar{x},2\bar{t}+1)+U_{\hat{t}}(2\bar{x}+1,2\bar{t})
\right) \rightarrow \mbox{max} 
\end{equation}
\item
Construct the link between two sites of the coarse lattice as close as
possible to the links between the corresponding fine lattice blocks, i.e.,
\begin{eqnarray}
\label{eq:block}
\left(
(U^{\dag}_{\hat{t}}(2\bar{x},2\bar{t}+1)+U^{\dag}_{\hat{t}}
(2\bar{x}+1,2\bar{t}+1))
\bar{U}_{\hat{t}}(\bar{x},\bar{t})
\right) & \rightarrow & \mbox{max} \quad \mbox{and} \\
\left(
(U^{\dag}_{\hat{x}}(2\bar{x}+1,2\bar{t})+U^{\dag}_{\hat{x}}
(2\bar{x}+1,2\bar{t}+1))
\bar{U}_{\hat{x}}(\bar{x},\bar{t}) \nonumber
\right) & \rightarrow & \mbox{max} 
\end{eqnarray}
\end{itemize}

Figure~\ref{fig:block} gives a picture of how the blocking is done.

As stated above, we calculated the fermion determinant on the blocked
lattice and compared it to the determinant on the full lattice. As shown
in Fig.~\ref{fig:det}, the blocked determinant follows the full
determinant closely.  Moreover the eigenvalues of $V$ exactly
equal to $-1$ are always preserved by the blocking.
Thus one would expect the effects of
dynamical fermions to be well reproduced by an approximation where
the blocked determinant is used instead of the full determinant. 

\begin{figure}[!ht]
\begin{center}
\includegraphics[height=8cm,width=10cm]{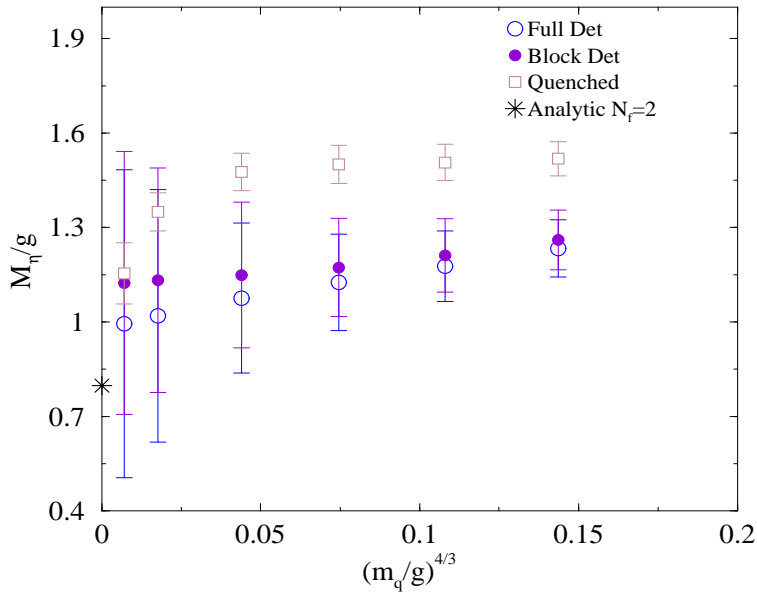}
\caption{\label{fig:etaNf2_grid}
\it{$M_\eta/g$ vs.~$(m/g)^{4/3}$ 
with the full, blocked and quenched 
determinant for the full operator for $N_f=2$.}}
\end{center}
\end{figure} 

This is confirmed by results shown in Figs.~\ref{fig:mpiNf2_grid}
and~\ref{fig:etaNf2_grid}.  In Fig.~\ref{fig:mpiNf2_grid} we
reproduce the values of the isotriplet mass $M_\pi$ obtained by using
either the full determinant or the blocked determinant in the
calculation of the observables.  We see that the results obtained with
the blocked determinant are in very good agreement with those obtained
with the full determinant. The physical results are those for
$N_f=2$.  We inserted in the figure also the results of a quenched
calculation and of a calculation with $N_f=1$ in order to highlight
the effects of the determinant.  
In particular, one sees that the data support a linear dependence
of $M_\pi$ versus $(m_q/g)^{2/3}$ with $M_\pi$ vanishing 
in the chiral limit only when one has $2$ dynamical quark flavors, 
i.e.~when one uses the square of the full determinant
${\rm Det}^2 (D)$ as weight factor. (This observation is based
on the data for the 4 rightmost points in the figure, which have
$N_x M_{\pi} < 4$ and thus are less likely to be affected by
finite size effects -- see also the fit in Fig.~\ref{fig:mpiNf2}
and the discussion that follows.)  This nontrivial result does not
change however, when ${\rm Det}^2 (D)$ is replaced by the square of
the blocked determinant ${\rm Det}^2_{\rm block} (D)$.

We reproduce in Fig.~\ref{fig:etaNf2_grid} the results for the
singlet mass $m_\eta$ for $N_f=2$. Once again we also include
the results of a quenched calculation to show the effect of the
determinant.  In this case also unquenching with the blocked
determinant is consistent with unquenching with the unblocked
determinant. The results however are less impressive, since the 
observable contains disconnected propagators and is therefore very noisy. 

\section{Conclusions}
\label{sec:conclusions}

We have explored, in the context of the overlap formulation of lattice
fermions, a scheme of approximation which focuses on the physical
properties of the system and takes advantage of the special features
of the Neuberger-Dirac operator.  Our method of approximation is based
on the projection of the Wilson operator on a subspace of lower
dimensionality, where the Neuberger-Dirac operator and its inverse are
constructed by dense matrix techniques.  We tested our approximation
on the Schwinger model, finding satisfactory results.  Of course, the
main interest of any scheme of approximation for overlap fermions is
in its applicability to four-dimensional QCD.  The possibility of a
successful extension of our method to this theory will depend on how
much one can reduce the dimensionality of the space used for the
projection with respect to the full space, while keeping a good
approximation to the physical observables.  

In the two-dimensional Schwinger model, the simplest implementation of these
ideas has been able to reduce the dimensionality of the fermionic vector
space by a factor of approximately 4. If this is an indication that one can
achieve a reduction of a factor of two per space-time dimension, it would
mean that in four dimensions the dimensionality of the fermionic space can be
reduced by a factor of order 16.  This may not be enough for QCD calculations
on lattices of realistic size, but even larger reductions of dimensionality
are not necessarily out of the question.  In our experiment with the
Schwinger model, we found that we could reduce the dimensionality of the
subspace used for the projection by a factor ranging well above 4 (as much as
10 or more while still keeping a good approximation to physical observables)
for the configurations with trivial topology.  As mentioned in
Section~\ref{sec:approx}, for configurations with non-trivial topology the
infrared properties of the Wilson operator are well captured even on small
subspaces, but the Neuberger projection mixes then eigenvectors with
different chiralities. This could however be more a shortcoming of the basis
used for the projection, i.e.~the Fourier basis, than of the method of
approximation in itself.

Even if it turned out that the reduction of dimensionality one can
achieve is not sufficient for a practical use of dense matrix techniques
in the projection subspace, the approximation we have studied can
still be of value.  The Neuberger-Dirac operator and its inverse
could be calculated within the subspace with sparse matrix numerical 
techniques similar to those currently in use for the whole space,
with the advantage of having to deal with a substantially smaller
system.  Alternatively, the approximate eigenvectors obtained by
the projection could be used for a preconditioning of the calculation
of the propagators on the full space.  This is a possibility which
we have not explored in our work, mostly for limitations of time and 
resources, but which might be worth studying in a future investigation.
It is likely, though, that incorporating in the computational techniques
used for dealing with overlap fermions as much insight as possible  
on the properties of the physical excitations will pay handsome
rewards, and we are currently investigating the application of
our approximation method to four-dimensional lattice QCD. 

\section*{Acknowledgments}
We wish to thank S.~Capitani, H.~Neuberger and U.~Heller for 
interesting discussions. This work has been supported in part under 
DOE grant DE-FG02-91ER40676. 
   
\end{document}